\documentclass[twoside]{article}
\usepackage{fleqn,espcrc2,epsfig}

\usepackage{graphicx}
\usepackage[figuresright]{rotating}

\title{
\hfill\begin{minipage}{0pt}\scriptsize \begin{tabbing}
	\hspace*{\fill} Edinburgh 2001/12\\ \end{tabbing}\end{minipage}\\[8pt]
	\vspace{-1.0cm}
The Isgur-Wise function on the lattice}
\author{UKQCD collaboration\\
	Presented by G.~N.~Lacagnina\address{Department of Physics and 
	Astronomy,  University of Edinburgh, Edinburgh EH9 3JZ, UK}} 
\begin{document}

\begin{abstract}
The $h_+, h_{A_1}$ form factors for the semi-leptonic $B\to D$ and
$B\to D^*$ decays are evaluated in quenched lattice QCD with
$\beta=6.2$. The action and the operators are fully $\mathcal{O}(a)$
non-perturbatively improved. The Isgur-Wise function is evaluated and
fitted to extract its slope; the latter is found to be
$\rho^2=1.1(2)(3)$ from the $B\to D^*$ decay and $\rho^2=1.0(2)(3)$
from the $B\to D$ decay. The form factors ratios $R_1,\ R_2$ are
evaluated and found to be in agreement with experimental
determinations.
\end{abstract}

\maketitle

\section{Introduction}
The $B\to D$ and $B\to D^*$ semi-leptonic decays have a considerable
phenomenological interest, since they can be studied to determine the
modulus of CKM matrix element $V_{\rm cb}$. Furthermore, the presence
of non-perturbative physics in these decays makes a lattice study
particularly appealing. The predictions of the Heavy Quark Effective
Theory (HQET) can be used to constrain the form of the matrix elements
that describe semi-leptonic decays of heavy-light mesons. In
particular, the relevant matrix elements are expressed in terms of a
set of form factors, that contain the non-perturbative physics of the
decay:

\begin{eqnarray}
\frac{\langle D(v')|{\bar c}\gamma^\mu b|B(v)\rangle}{\sqrt{m_B m_D}} & = & \nonumber\\(v+v')^\mu h_{+}(\omega)+(v-v')^\mu h_{-}(\omega)\ ,\\
\frac{\langle D^*(v',\epsilon)|{\bar c}\gamma^\mu b|B(v)\rangle}{\sqrt{m_B m_{D^*}}} & = &\nonumber\\
 i\epsilon^{\mu\nu\rho\sigma}\epsilon_{\nu}^{*}v_{\rho}'v_{\sigma}h_V(\omega)\ ,\\ 
\frac{\langle D^*(v',\epsilon)|{\bar c}\gamma^\mu\gamma^{5} b|B(v)\rangle}{\sqrt{m_B m_{D^
*}}} & = &\nonumber\\ (\omega+1)\epsilon^{*\mu}h_{A_1}(\omega)&+&\nonumber \\ -[h_{A_2}(\omega)v^\mu+h_{A_3}(\omega)v'^{\mu}](\epsilon^*\cdot v)
\end{eqnarray} 

where $v,\ v'$ are the velocities of the initial and the final meson
respectively, and $\omega=v\cdot v'$; $\epsilon$ is the polarisation
vector of the $D^*$ meson. In the heavy quark limit, the six form
factors become related to a universal function known as the Isgur-Wise
function, $\xi(\omega)$. However, there are two kinds of corrections
to the heavy quark symmetry that one has to take into account:
perturbative QCD corrections and symmetry breaking corrections
proportional to inverse powers of the heavy quark mass. In terms of
these corrections, the relations between the form factors and
$\xi(\omega)$ are \cite{neubert_1}:

\begin{eqnarray}
h_{j}(\omega)&=&[\alpha_j+\beta_j(m_b,m_c;\omega)+\nonumber\ \\
&&\gamma_j(m_b,m_c;\omega)+...]\xi(\omega)\ ,
\end{eqnarray}

The $\alpha_j$ terms are constants that fix the
behaviour of the form factors in absence of corrections ($\alpha_{+} =
\alpha_V =\alpha_1 = \alpha_3 = 1$; $\alpha_-=\alpha_2=0$).  The $\beta_j$ and
$\gamma_j$ functions account for radiative corrections and power
corrections of ${\mathcal O}(1/m_{b,c})$ respectively. The radiative
corrections are calculable in perturbation theory, while the power
corrections are non-perturbative in nature \cite{neubert_2}. At zero
recoil ($v=v',\omega=1$), however, Luke's theorem \cite{luke}
guarantees that both $\gamma_+$ and $\gamma_{A_1}$ are of order
${\mathcal O}(1/m_{b,c}^2)$.  At small recoil, the Isgur-Wise function
is modeled by:

\begin{equation}
\label{eqn:linearwisgur}
\xi(\omega)=1-\rho^2(\omega-1)+{\mathcal O}((\omega-1)^2)\ ,
\end{equation}

where $\rho^2$ is called the slope parameter and $\xi(1)=1$ because of
current conservation. Alternative parametrisations of $\xi(\omega)$
have been proposed, which start to differ from
(\ref{eqn:linearwisgur}) at ${\mathcal O}((\omega-1)^2)$ \cite{models_1,models_2}.

\section{Simulation details}

The calculations presented in this work were performed using a
non-perturbatively ${\mathcal O}(a)$ improved action (SW, $c_{\rm
SW}=1.614$), within the quenched approximation, at $\beta=6.2$, on a
set of $216$ configurations, with a volume of $24^3\times 48$. The
improvement coefficients for the current operators were taken from the
work of Bhattacharya {\em et al.} \cite{bhattacharya}.  The inverse
lattice spacing, set using the Sommer scale $r_0$ \cite{sommer_scale},
was $a^{-1}\simeq 2.9\ ({\rm GeV})$. The matrix elements have been
extracted for eight combinations of the heavy quark masses ($m_Q\simeq
m_{\rm charm}$) and two values of the mass of light (passive) quark
($m_q\simeq m_{\rm strange}$). Since, at fixed $\omega$, the residual
dependence of $\xi(\omega)$ on the heavy quark mass was statistically
negligible, no extrapolations were necessary.

\section{Extraction of the Isgur-Wise function}

The Isgur-Wise function has been extracted independently from the
$h_+$ and $h_{A_1}$ form factors, that at $\omega=1$ are protected
from ${\mathcal O}(1/m_{\rm b,c})$ power corrections by Luke's theorem:

\begin{eqnarray}
\label{eqn:wisgur}
\xi(\omega) & \simeq & \frac{h_+(\omega)}{1+\beta_+(\omega)}\ ,\\
\xi(\omega) & \simeq & \frac{h_{A_1}(\omega)}{1+\beta_{A_1}(\omega)}
\end{eqnarray}

\begin{figure}
\vspace{-2.5cm}
\epsfig{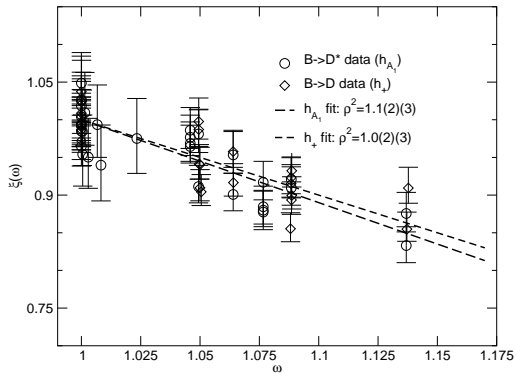}
\caption{The Isgur-Wise function from the $B\to D$ and $B\to D^*$ semi-leptonic decays.}
\label{fig:wisgur}
\end{figure}

The power corrections have been neglected, as they have been estimated
and found to be consistent with zero in the range $1.0\le \omega \le
1.2$.  It is not possible to extract the Isgur-Wise function from the
form factors for which $\alpha_j=0$ as they are just a collection of
QCD corrections; similarly, the $h_V$ form factor could not be used
because its power correction has been found large.  The Isgur-Wise
function has been fitted to the linear model
(\ref{eqn:linearwisgur}). Unconstrained fits have also been performed,
i.e. relaxing the condition $\xi(1)=1$. In both decays, $\xi(1)$ was
found to be consistent with one. Results are summarized in Table
(\ref{tab:wisgur_fits}) and plotted in Figure (\ref{fig:wisgur}). The
data from the two decays are in excellent agreement with each other
and lie consistently on the same line.

\begin{table}
\begin{center}\caption{Results of the linear fits the Isgur-Wise function,
	with and without the $\xi(1)=1$ constraint. Quoted errors are
	statistical.\smallskip}
\label{tab:wisgur_fits}
\begin{tabular}{@{}llcccc}
\hline
&& \multicolumn{2}{c}{$B\to D$}
 &\multicolumn{2}{c}{$B\to D^*$}
\\
&&con. fit&un. fit&con. fit&un. fit\\
\hline
&$\xi(1)$ & $\equiv 1$ &$0.99(2)$ &$\equiv 1$  &$1.0(3)$\\
   &$\rho^2$ & $1.0(2)$ &$0.9(3)$ &$1.1(2)$ &$1.1(2)$\\
\hline
\end{tabular}
\end{center}
\end{table}

\section{Ratios of form factors}

The two following ratios of form factors have also been calculated:

\begin{eqnarray}
R_1(\omega)&=&\frac{h_V(\omega)}{h_{A_1}(\omega)}\ , \\
R_{2}(\omega)&=&\frac{h_{A_3}(\omega)+\frac{M_{\rm D*}}{M_{\rm B}}h_{A_2}(\omega)}{h_{A_1}(\omega)}
\end{eqnarray}

\begin{figure}
\epsfig{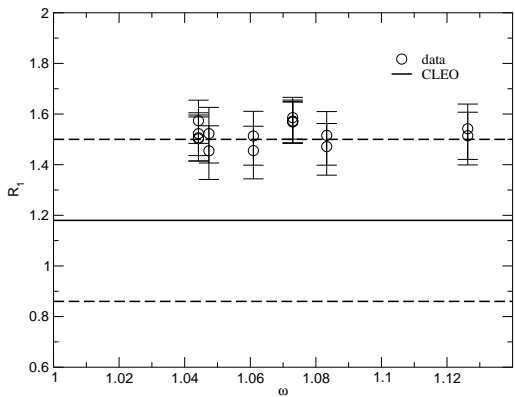}
\caption{The $R_1$ ratio. The range of the experimental determination is shown.}
\label{fig:R1}
\end{figure}

\begin{figure}
\vspace{-1.0cm}
\epsfig{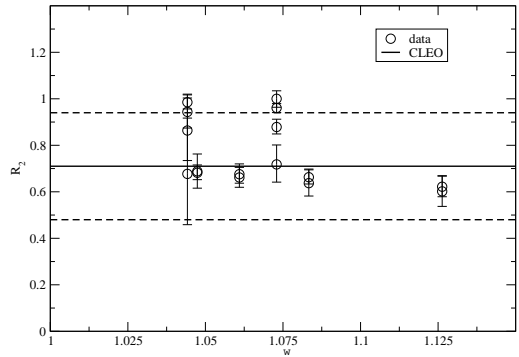}
\caption{The $R_2$ ratio. The range of the experimental determination
is shown.}
\label{fig:R2}
\end{figure}

These two ratios would be equal to one in the absence of symmetry
breaking corrections. Figures (\ref{fig:R1}) and (\ref{fig:R2}) show
that this work's determinations are in agreement with the experimental
determinations by the CLEO collaboration \cite{cleo_ratios}. 

\section{Systematic uncertainties}

The systematic uncertainty on the slope of the Isgur-Wise function has
been estimated considering the effect of the choice of the
scale-fixing quantity, the residual light quark mass dependence, and
the effect of using different non-zero momentum kinematic channels.
Three alternative models of the Isgur-Wise function have been
considered, and the variation in the extracted slope parameter has
been taken as a systematic uncertainty.

\section{Summary and conclusions}

The slope of the Isgur-Wise function has been extracted from a lattice
study of the $B\to D$ and $B\to D^*$ semi-leptonic decays, using a
${\mathcal O}(a)$ non-perturbatively improved action at
$\beta=6.2$. The results from the two decays have been found in good
agreement: $\rho^2=1.1(2)(3)$ from $B\to D^*$ and $\rho^2=1.0(2)(3)$
from $B\to D$. This result is in good agreement with previous
calculations performed by members of the UKQCD collaboration, which
did not use a full set of non-perturbatively determined improvement
coefficients \cite{prev_ukqcd,ukqcd_2,ukqcd_3,ukqcd_4}. The ratios of
form factors $R_1,R_2$ have been calculated and found in agreement
with experimental data.

\section{Acknowledgements}
Work supprted by the European Community's Human potential programme
under HPRN-CT-2000-00145 Hadrons/LatticeQCD and PPARC grant 
${\rm PPA}/{\rm G}/{\rm O}/1998/00621$.

\end{document}